\documentclass{kapproc} 

  \usepackage{procps}
\usepackage{graphicx}

\normallatexbib

\begin{document}

\articletitle{Superconducting Quantum Computing \ \\ without
Switches}


\author{Marc J. Feldman and Xingxiang Zhou}
\affil{Department of Electrical and Computer Engineering\\
University of Rochester\\
Rochester, NY 14627, USA}

\email{feldman@ece.rochester.edu}


\email{xizhou@ece.rochester.edu}

\begin{abstract}
This paper presents a very simple architecture for a large-scale
superconducting quantum computer.  All of the SQUID qubits are
fixed-coupled to a single large superconducting loop.
\end{abstract}


\section{Introduction}
What physical system is most appropriate for building a quantum
computer? The proponents of one implementation or another usually
discuss this question by reference to DiVincenzo's criteria
\cite{DiVincenzo2000}. These five criteria very nicely summarize
the requirements for the physical realization of a quantum
information processing system.

For our purposes, four of these criteria can be combined into a
single one:  good qubits. That has many implications, but we will
say no more about it here. For this paper, let us assume ideal
qubits. The other criterion, number four, is ``A `universal' set
of quantum gates''. This addresses the interactions between qubits
rather than the nature of the qubits themselves. A quantum
computer is, at least, a set of interacting qubits.

In this paper we will first discuss the ``no switch'' problem
regarding superconducting implementations of quantum computing. In
brief, it has been difficult to come up with a satisfactory scheme
to switch the coupling between two superconducting qubits on and
off. Then we will mention several possibilities for quantum
computing using fixed, rather than switchable, couplings between
qubits, and indicate why these are unsatisfactory for
superconducting qubits as well.

Our solution to this situation is based on recent work
\cite{Zhou2002} in which a virtual switch, rather than a
substantial physical switch, is realized by carrying out the steps
of the quantum computation in and out of designed ``interaction
free subspaces'' which are analogous to decoherence free
subspaces. We will give examples of how these virtual switches can
be employed in a variety of different architectures for a
superconducting quantum computer.

Finally we explore one particular architecture at much greater
length. Many SQUID qubits are fixed-coupled to a single large
superconducting loop. We show that this is adequate for
large-scale quantum computing, and specify the requisite
parameters. The parameters are chosen for rf-SQUID qubits
\cite{Bocko1997,Lukens2000} but this architecture is equally
appropriate for persistent current SQUID qubits \cite{Mooij2000}
as well.

\section{The ``No-Switch'' Problem}
Quantum algorithms are generally formulated in terms of a
collection of qubits subject to a sequence of single-qubit
operations and two-qubit gates. A two-qubit gate such as the CNOT
can be represented by a unitary $4\times 4$ matrix over the bases
of the two qubits. Taken literally, such an algorithm implies that
there are three distinct modes of operation of a quantum computer.
Two qubits have: 1) the idle mode in which information is stored
in qubits which do not evolve, 2) the single-qubit operation mode
in which local fields applied to qubit 1 have no effect on qubit
2, and 3) the two qubit operation mode in which qubit 1 and qubit
2 are coupled together and a quantum gate is realized through the
coupling Hamiltonian. The two-qubit Hamiltonian that expresses
this is
\begin{equation}
H=- \vec{B}_1(t)\cdot \vec{\sigma}_1- \vec{B}_2(t)\cdot
\vec{\sigma}_2 +\sum_{\alpha,\beta}J_{12}^{\alpha
\beta}(t)\sigma_1^\alpha\sigma_2^\beta,
\end{equation}
where $\sigma$'s are the Pauli matrices, $\vec{B}_i$ is the local
field at qubit $i$, and $J_{12}$ is the coupling strength.
$\vec{B}_i$ and $J_{12}$ are time dependent under external
control. To alternate between the three operational modes it is
necessary that $\vec{B}_i$ be turned on and off as required and
$J_{12}$ be turned on and off as required.  In other words, there
must be a switch between qubit 1 and qubit 2.

\begin{figure}[h]
    \centering
    \includegraphics[width=3.8in, height=1in]{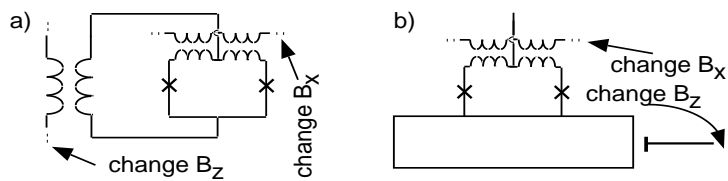}
    \caption{(a) A flux qubit
biased near $\Phi_0/2$, for which the remove from one-half flux
quantum acts as the $B_z$ field, and the suppression of the
effective critical current acts as the $B_x$ field. (b) A charge
qubit, for which the remove from a single-charge voltage bias acts
as the $B_z$ field, and the suppression of the effective critical
current acts as the $B_x$ field.}
    \label{fig:BxBz}
\end{figure}

\begin{figure}[h]
    \centering
    \includegraphics[width=3.8in, height=0.7in]{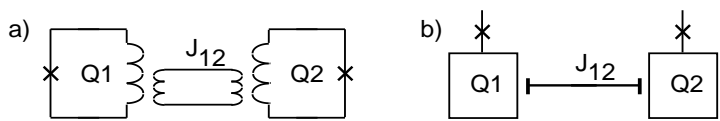}
    \caption{(a) Inductively coupled flux qubits. (b) Capacitively coupled
    charge qubits.}
    \label{fig:J12}
\end{figure}

In both flux and charge \cite{Nakamura1999,Vion2002}
superconducting qubits, the control of the time dependence of
$B_x$ and $B_z$ is relatively ``easy''. 
As illustrated in Fig. \ref{fig:BxBz}, time dependent $B_x$ and
$B_z$ fields on superconducting qubits are achieved by simply
varying the biases. The fixed coupling between qubits is ``easy''
for the superconducting qubits as well, as shown in Figure
\ref{fig:J12}. 
The flux qubits can be coupled by a simple inductive connection,
and the charge qubits can be coupled by a simple capacitive
connection between them. This is certainly ``easy'' too.  In fact
a fixed inductive coupling between rf-SQUID qubits was diagrammed
in the first paper written on superconducting quantum computing
\cite{Bocko1997}.

Variable coupling between superconducting qubits is much harder.
There are many proposed schemes, but it is generally agreed that
none of these proposed switches is completely satisfactory
\cite{Makhlin2001}. Some early examples are Averin's proposal
\cite{Averin1998} to couple charge qubits by moving the charge
among the charge islands, and Sch\"{o}n {\it et al.}'s LC
resonator ``bus'' \cite{Makhlin2001,Makhlin1999}. Perhaps the
simplest variable superconducting switch was first described by
Mooij {\it et al.} \cite{Mooij1999}. It is essentially a solid
superconducting loop broken by a Josephson junction which can be
open circuited - it is an rf SQUID in which the critical current
can be reduced to zero. More recent schemes are a differential
version of the rf-SQUID switch \cite{Filippov2002}, the ``Bloch
transistor'' \cite{Likharev2002}, and the INSQUID
\cite{Clarke2002}.

These are only a few of the proposals for switches linking the
qubits in a superconducting quantum computer. Some of these
schemes are not practical, but for the most part they could
possibly be used but they just seem awkward. Many of them require
external controls, which are likely to be important sources of
decoherence. Many require an unrealistic level of parameter
control to operate successfully. Others appear to be difficult to
scale to a many qubit system. In any case, none of these proposals
has garnered outside support, and new and more elaborate proposals
keep appearing in the literature.

\section{Fixed-Coupled Quantum Computer}
One may consider making a quantum computer without switches.  Any
interaction between qubits, fixed or variable, is sufficient for a
universal quantum gate. Such a computer however may be more
difficult to realize in practice.

An early example is the spin-lattice quantum computer
\cite{Lloyd1993}, in which the ``spins'' (qubits) are hard-wired
into a lattice. There are no switches; rather, the entire lattice
is addressed by global fields. This paradigm for quantum computing
requires more complex manipulations to perform simple operations,
and it can be tedious to render quantum algorithms into lattice
interactions. Although widely referenced, this scheme has not been
much adopted by others. For superconducting systems, the spin
lattice architecture seems a particular waste of resources, in
that single-bit operations which are``easy'' for superconducting
qubits are not at all utilized.

A prominent quantum computer architecture today is the NMR
molecular system. In NMR, coupling between the qubits is indeed
fixed. There is no reason in principle not to build a hard-wired
superconducting quantum computer following the NMR model. In the
NMR model, complex synthesized ``refocusing pulses'' are required
to reverse the evolution of unwanted phase-shifts incurred by the
always-on couplings. The complexity of such refocusing pulses
grows with the size of the system. More discouraging, the
refocusing pulses for a superconducting system would need be at
many orders of magnitude higher frequency than for NMR, and this
may be impossible to achieve with the precision required, with
today's technology.

So it is seen that switches between qubits are not absolutely
essential for quantum computing, but they are likely to be a
practical necessity for large qubit systems.

\section{Interaction Free Subspaces}

Recent work \cite{Zhou2002} has shown that logical qubits
consisting of two or more physical qubits can be constructed to
code quantum information in an ``interaction free subspace'' (IFS)
such that there is no interaction between these qubits even though
they are physically coupled.  This is analogous to the more
familiar ``decoherence free subspace'' (DFS), which can be
employed to isolate quantum information from interacting with
environmental modes which would lead to decoherence
\cite{Zanardi1997}. The DFS concept is widely utilized; it assumes
that there are symmetries in the coupling of the qubits to their
environment, and employs those symmetries to avoid decoherence.
The IFS is different in that it relies on symmetries in the
coupling between qubits which can be created by the experimenter.
Information is coded in such a way that when a logical qubit is in
its IFS it is not affected by other qubits it
is physically coupled to ({\it i.e.}, switch is open). 
Single bit operations can be performed when the neighboring qubits
are in their IFS. When two coupled logical qubits are removed from
their IFS, two bit gates can be performed ({\it i.e.}, switch is
closed). Other logical qubits in their IFS are not affected by
these operations.

\begin{figure}[h]
    \centering
    \includegraphics[width=3.5in, height=1.1in]{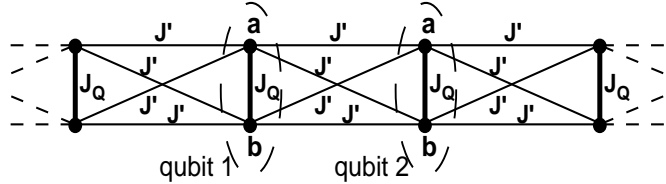}
    \caption{Example quantum computer Architecture.
    Each dot is a physical qubit and the lines represent couplings.
    Two qubits (a, b) connected by a vertical line is an encoded qubit.
    }
    \label{fig:linearQC}
\end{figure}

IFS operations are discussed at length in \cite{Zhou2002}.  Here
we will merely present an example. We assume diagonal interactions
between physical qubits of the form $J_{12}\sigma_1^z\sigma_2^z$.
Our quantum computer is a one-dimensional Ising lattice of logical
IFS qubits as illustrated in Fig. \ref{fig:linearQC}. For the
Ising interaction it is sufficient to use two physical qubits, the
dots labelled `a' and `b' in the Figure, to compose the logical
qubit, ``qubit 1''. The coupling between the physical qubits are
either $J_Q$ or $J'$ as labelled.  Then it is easy to see that the
following basis states
\begin{equation}
|0\rangle=|\uparrow_a,\downarrow_b\rangle,
|1\rangle=|\downarrow_a,\uparrow_b\rangle
\end{equation}
are annihilated by the interaction Hamiltonian
$H_{int}=J'(\sigma_{1a}^z+\sigma_{1b}^z)(\sigma_{2a}^z+\sigma_{2b}^z).$
Therefore, if these states are used to code quantum information,
the logical qubits do not affect each other.

Arbitrary single bit operations can be performed on logical qubit
1 using $\vec{B}_a(t)$ and $\vec{B}_b(t)$ with the fixed a - b
coupling $J_Q$. Neighboring logical qubits must remain each in its
IFS during the operations on qubit 1.  Reference \cite{Zhou2002}
details how a CNOT gate can be performed on qubit 1 and qubit 2.
The CPHASE gate (equivalent to CNOT up to single bit gates) is
achieved by: first flip the state of both qubit $1b$ and qubit
$2b$ to remove the logical qubits from IFS; then follow a set of
prescribed rotations of the physical qubits and allow the logical
qubits to interact for a certain amount of time; then again flip
the state of both qubit $1b$ and qubit $2b$ to return the logical
qubits back into IFS. The time required for this is $\pi/16J'$.

The linear architecture shown in Fig. \ref{fig:linearQC} is one of
many possibilities, chosen for clarity of explanation alone. It is
equally as possible to make a two-dimensional array. In the next
section we will consider in detail a very different architecture
in which all of the physical qubits are inductively fixed-coupled
to a single large superconducting loop. Notice that this prevents
the use of parallel operations because only a few qubits can be
out of their IFS at once. The linear array would allow great
parallelism if two-thirds of all qubits are concurrently
undergoing two-qubit gate operations. Between these two extremes
is an architecture with qubit clusters coupled together by link
qubits. It is seen that the IFS virtual switch encourages great
versatility in quantum computer architecture. Eventually, one may
hope that the computer architecture be designed for greatest
suitability for classes of quantum algorithms to be addressed.

\section{The Inductor Bus Quantum Computer}
Figure \ref{fig:linearQC} seems to imply that close-coupled
physical qubits should compose a logical qubit, and this logical
qubit is coupled to only several other logical qubits. In fact,
neither of these constraints is necessary. An architecture in
which every physical qubit is coupled to every other physical
qubit, with all equal coupling strength, satisfies the conditions
for IFS as well. There are very natural implementations for this
kind of quantum computer architecture using superconducting
qubits. For instance, many superconductor charge qubits could be
capacitively coupled to a single floating conductor island. We
will examine the situation where many SQUID qubits are coupled
each by a fixed mutual inductance to a single large solid
superconducting loop -- the ``inductor bus''.

\begin{figure}[h]
    \centering
    \includegraphics[width=4.5in, height=2.8in]{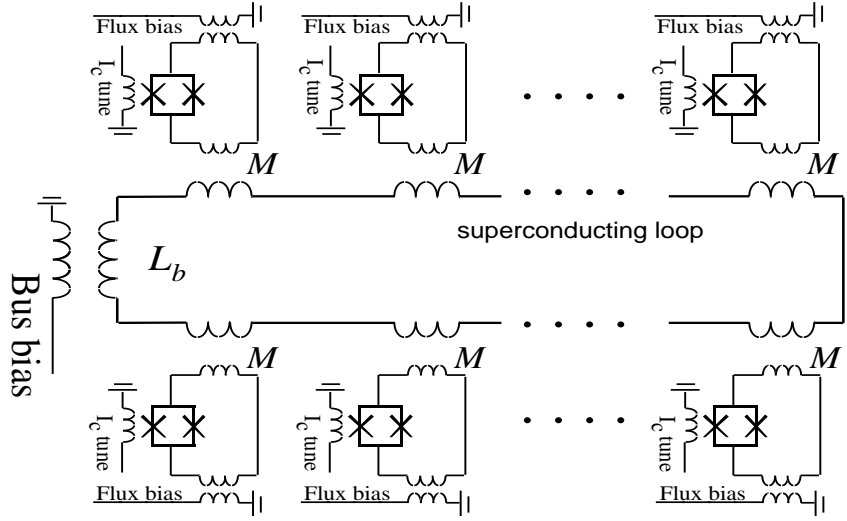}
    \caption{The ``inductor bus'' quantum computer.}
    \label{fig:loopQC}
\end{figure}

The inductor bus quantum computer is illustrated in Fig.
\ref{fig:loopQC}. $N$ (an even number) identical rf-SQUIDS are
inductively coupled to a superconducting inductor loop with self
inductance $L_b$ ($b$ stands for ``bus''). All couplings have the
same mutual inductance $M$. The flux linking the bus inductor loop
is $\Phi_b$. Its external flux bias is $\Phi_{bx}$. The current in
the bus is $I_b$. All rf-SQUIDS have the same inductance $L$, the
same capacitance $C$ and the same Josephson energy $E_J
(=I_c\Phi_0/2\pi)$. The total flux, the flux bias and the current
of the $i$th ($i=1,2...N$) rf-SQUID are $\Phi_i$, $\Phi_{ix}$ and
$I_i$. Note that $\Phi_b$, the total magnetic flux in the loop
cannot change because it is a solid superconducting loop. So when
the current in one of the rf-SQUIDS changes the current in the
loop must change slightly to maintain $\Phi_b$.  This couples to
all of the other rf-SQUIDS.

\subsection{The System Hamiltonian}
We mentioned that the flux in the inductor loop must be quantized
in units of the flux quantum $h/2e$: $\Phi_b=n\Phi_0$, $n$ being
any integer. For simplicity let us take $n = 0$ and $\Phi_{bx}=0$.
We can write down the flux equations for the rf-SQUIDS and the
bus, taking into account all biases on the inductive loops:
\begin{eqnarray}
\Phi_i=\Phi_{ix}+MI_b+LI_i, i=1,2...N, \label{Phi_i}\\
L_bI_b+\sum_{i=1}^N MI_i=0. \label{Phi_b2}
\end{eqnarray}
These equations allow us to solve for the currents in terms of the
fluxes, which can be used to calculate the inductive energy
$E_{ind}=\frac{1}{2}\sum_{i=1}^N LI_i^2 +\frac{1}{2} L_bI_b^2
+\sum_{i=1}^N MI_bI_i$ of the system. We are interested in the
limit of weak coupling,
\begin{equation}
NM^2/LL_b\ll 1. \label{weak}
\end{equation}
Keeping terms to lowest order in $M^2/LL_b$, and adding the
charging and Josephson energies of the rf-SQUIDS, we obtain the
system Hamiltonian
\begin{equation}
H=\sum_{i=1}^N H_i+H_{int}, \label{Hami}
\end{equation}
where $H_i$'s are the self Hamiltonians of the rf-SQUIDS (with the
normalized self inductance),
\begin{equation}
H_i=-\frac{\hbar^2}{2C}\frac{\partial^2}{\partial\Phi_i^2}
+\frac{(\Phi_i-\Phi_{ix})^2}{2L/(1+\frac{M^2}{LL_b})}
-E_J\cos(2\pi\frac{\Phi_i}{\Phi_0}),
\end{equation}
and $H_{int}$ is the interaction Hamiltonian between pairs of
rf-SQUIDS,
\begin{equation}
H_{int}=\frac{M^2}{L_b}\sum_{i>j}^N
\frac{(\Phi_i-\Phi_{ix})}{L}\frac{(\Phi_j-\Phi_{jx})}{L}.
\label{intH}
\end{equation}
The summation in Eq. (\ref{intH}) is between each pair of
rf-SQUIDS. 
Eq. (\ref{intH}) describes the coupling between each pair of
isolated rf-SQUIDS through an effective mutual inductance $M_{eff}
=M^2/L_b$. The other effect seen here is that the self inductance
of the rf-SQUIDS is renormalized slightly:
\begin{equation}
L\rightarrow L/(1+M^2/LL_b).
\end{equation}
Under low temperatures, if we bias the rf-SQUIDS appropriately
they are effectively two state systems. Then we can rewrite Eq.
(\ref{Hami}) using the Pauli matrices,
\begin{equation}
H=\sum_{i=1}^N \{-\hbar\Delta_i(I_i^c)\sigma_i^x+
\varepsilon_i(\Phi_{ix})\sigma_i^z \} +\sum_{i>j}^N J \sigma_i^z
\sigma_j^z, \label{H}
\end{equation}
where $\Delta_i$ and $\varepsilon_i$ are the tunneling matrix
element and energy offset between the two potential wells of the
$i$th rf-SQUID, controllable via its critical current $I_i^c$ and
flux bias $\Phi_{ix}$. The eigenstates of $\sigma^z$ correspond to
the left and right well localized states. $J$ is the always on and
fixed coupling strength between the rf-SQUIDS proportional to the
effective
mutual inductance $M^2/L_b$. 

\subsection{Initialization and computation}
This has been discussed in Ref. \cite{Zhou2002}. 
The first step is to flux bias all the rf-SQUIDS near to the
symmetrical point $\Phi_0/2$, but far enough away to assure that
they end up in the lower of their two flux localized states. Then
raise the barrier height of the rf-SQUID potential such that there
is no tunneling between the two flux states ($\Delta=0$), and turn
their flux bias to the symmetry point $\Phi_0/2$. The rf-SQUIDS
are left in the left (or right) well. At the symmetry point
$\varepsilon=0$ as well, therefore the Hamiltonian of the qubits
is 0, and the qubit state is frozen.

Two of the physical qubits are chosen to constitute a logical
qubit. (The choice is arbitrary!) The next step is to flip the
state of one of these physical qubits. This can be done by
lowering its potential barrier to obtain a finite tunneling rate
$\Delta$ and letting it evolve for $t = \pi/\hbar\Delta$ (a $\pi$
pulse). The qubit will shift from the left well to the right well.
The potential barrier is restored to freeze the state of the
qubit. Now the two physical qubits are in the IFS
($|\uparrow\downarrow\rangle$ and $|\downarrow\uparrow\rangle$)
and the logical qubit is decoupled from the bus. In their IFS, the
two physical qubits apply a 0 net flux on the bus. The other
qubits are still in their left wells and they do apply a flux to
the bus. 
Then, all of the other qubits are treated the same way until all
logical qubits are driven into the IFS.

The large superconducting loop may start with flux in it, $n\neq
0$. Even if $n=0$, during the initialization there is a large
current in the loop and some current $I_b\neq 0$ will remain
because of inevitable variations in SQUID parameters.  These
currents have no effect on the IFS code states.  Still, one is
uncomfortable that, for large $N$, small unforseen errors might
compound.  If this is a concern, it is possible to keep $I_b=0$ by
changing the bus flux bias $\Phi_{bx}$ a little.  Another
possibility is to break the solid superconducting loop by a very
small series resistor, sub-$\mu\Omega$, such that $L_b/R$ is
comparable to the initialization time but much longer than the run
time of the quantum computer.

The computation proceeds through a series of single physical qubit
operations that induce rotations around the x and z directions of
each physical qubit.  These operations are realized by changing
the tunneling rate $\Delta$ and the energy offset $\varepsilon$ of
the rf-SQUIDS (cf. Eq. (\ref{H}) and Fig. \ref{fig:BxBz}a).  Then
a computation of any complexity can be performed using the logical
qubits in and out of the IFS, accessing only these single physical
qubit operations, as prescribed in Ref. \cite{Zhou2002}.

\subsection{Parameters}
Finally let us specify the parameters required for the inductor
bus quantum computer in Fig. \ref{fig:loopQC}. We will see that
the computer may include a large number of qubits, at least $N
\sim 1000$, for realistic parameters allowed by the current
technology, assuming as always ideal qubits.

The relevant parameters for a single rf-SQUID are $L$, $I_c$, and
$C$. We choose $L=150pH$, $C=80fF$, and $Ic=3 \mu A$. These are
familiar numbers, very similar to the values considered in
\cite{Feldman2001}. Solving the Schr\"{o}dinger equation
numerically, we find when the critical current is unsuppressed,
the tunneling matrix element $\Delta$ is about $30Hz$, therefore
the tunneling can be considered completely off in this case.
Suppressing $I_c$ down to $2.375 \mu A$ gives a tunneling matrix
element $\Delta\approx 2.6 GHz$, which allows to flip the state of
the rf-SQUID (a $\pi$ pulse in the $x$ direction) in about $0.4
ns$. Flux biasing the rf-SQUID off the symmetrical point
$\Phi_0/2$ by $0.15 m\Phi_0$ gives an $\varepsilon$ of about $2.7
GHz$. This allows rotations around the $z$ direction with a speed
of a few $GHz$.


The two-bit operation speed is determined by the strength of qubit
coupling, $J$. The single physical bit operations should be much
faster than the two bit operations. Referring to Eq. (\ref{H}),
this means that $\Delta$ and $\varepsilon$ (when they are on)
should be much larger than J (which is always on).
We choose the two-bit operation time to be tens of $ns$.
Evaluating the interaction Hamiltonian Eq. (\ref{intH}) in the
qubit bases, we find that an effective mutual inductance
($M^2/L_b$) of about $2fH$ results in a coupling strength $J$
of about $25MHz$. This is comfortably satisfied with $L_b=2nH$ and
$M=2pH$. This is satisfactory because $M \ll L$ and $M \ll L_b$,
so $M^2/LL_b \ll 1$. 


How many qubits $N$ can be attached to the bus? First of all, the
weak coupling limit (Eq. (\ref{weak})) requires
\begin{equation}
N \ll LL_b/M^2. \label{N}
\end{equation}
The other practical consideration is $N<L_b/M$, just by a simple
geometrical argument. 
We assume a planar circuit geometry where all inductors are
realized by single turn thin film conductors lithographed over a
ground plane
\cite{Feldman2001}. 
Then the inductance per unit length of conductor will be roughly
the same for all inductors, and $M \sim kL_{bi}$, where $L_{bi}$
is the section of $L_b$ coupled to the $i$th rf-SQUID, and $k$ is
the coupling constant, necessarily less than 1. Clearly, the
maximum $N = L_b/L_{bi} \sim kL_b/M$, as stated above. For $L_b=2
nH$ and $M = 2 pH$, $N_{max} = 1000$, which satisfies the weak
coupling limit Eq. (\ref{N}). $N$ in the inductor bus quantum
computer can be made larger than 1000 only by the undesirable
recourse of decreasing the coupling strength and the speed of two
bit operations.


\section{Conclusions}
We give a prescription for a large-scale $N \sim 1000$
superconducting quantum computer. It is based on the idea of
``interaction free subspace'' presented in \cite{Zhou2002}.
Solving the ``no switch'' problem without the need to use a
physical switch, 
it will help in the effort to construct a practical
superconducting quantum computer.


\end{document}